\def\Ni{\noindent}
\def\etal{{et\,al.}}

\def\msun{M$_{\odot}$}

\def\amin{\ifmmode ^{\prime}\else$^{\prime}$\fi}
\def\asec{\ifmmode ^{\prime\prime}\else$^{\prime\prime}$\fi}
\def\degs{\ifmmode ^{\circ}\else$^{\circ}$\fi}
\def\farcm{\hbox{$.\mkern-4mu^\prime$}}  
\def\h{$^{\rm h}$} \def\m{$^{\rm m}$} \def\s{$^{\rm s}$}
\def\grad{$^\circ$}

\def \SAIT #1 #2 {{\em Mem.\ Soc.\ Astron.\ It.\/} {\bf #1}, #2}
\def \MESS #1 #2 {{\em The Messenger\/} {\bf #1}, #2}
\def \ASTRNACH #1 #2 {{\em Astron. Nach.\/} {\bf #1}, #2}
\def \AAP #1 #2 {{\em Astron. Astrophys.\/} {\bf #1}, #2}
\def \AAL #1 #2 {{\em Astron. Astrophys. Lett.\/} {\bf #1}, L#2}
\def \AAR #1 #2 {{\em Astron. Astrophys. Rev.\/} {\bf #1}, #2}
\def \AAS #1 #2 {{\em Astron. Astrophys. Suppl. Ser.\/} {\bf #1}, #2}
\def \AJ #1 #2 {{\em Astron. J.\/} {\bf #1}, #2}
\def \ANNREV #1 #2 {{\em Ann. Rev. Astron. Astrophys.\/} {\bf #1}, #2}
\def \APJ #1 #2 {{\em Astrophys. J.\/} {\bf #1}, #2}
\def \APJL #1 #2 {{\em Astrophys. J. Lett.\/} {\bf #1}, L#2}
\def \APJS #1 #2 {{\em Astrophys. J. Suppl.\/} {\bf #1}, #2}
\def \APSS #1 #2 {{\em Astrophys. Space Sci.\/} {\bf #1}, #2}
\def \ASR #1 #2 {{\em Adv. Space Res.\/} {\bf #1}, #2}
\def \BAIC #1 #2 {{\em Bull. Astron. Inst. Czechosl.\/} {\bf #1}, #2}
\def \JSQRT #1 #2 {{\em J. Quant. Spectrosc. Radiat. Transfer\/} {\bf #1}, #2}
\def \MN #1 #2 {{\em Mon. Not. R. Astr. Soc.\/} {\bf #1}, #2}
\def \MEM #1 #2 {{\em Mem. R. Astr. Soc.\/} {\bf #1}, #2}
\def \PLR #1 #2 {{\em Phys. Lett. Rev.\/} {\bf #1}, #2}
\def \PASJ #1 #2 {{\em Publ. Astron. Soc. Japan\/} {\bf #1}, #2}
\def \PASP #1 #2 {{\em Publ. Astr. Soc. Pacific\/} {\bf #1}, #2}
\def \NAT #1 #2 {{\em Nature\/} {\bf #1}, #2}
\documentstyle[psfig]{memsait}
\begin{opening}
\title{Gamma-Ray Bursts: Old and New}
\author{JOCHEN GREINER}
\institute{Astrophysikalisches Institut Potsdam, 14482 Potsdam, Germany
   (jgreiner@aip.de)}
\date{} 
\end{opening}

\begin{document}

\oddpagefooter{}{}{} 
\evenpagefooter{}{}{} 
\ 
\bigskip

\begin{abstract}
Gamma-ray bursts  are sudden releases of energy that for a duration of
a few seconds outshine even huge galaxies.
30 years after the first detection of a gamma-ray burst  their origin
remains a mystery. Here I first review the ``old'' problems which have baffled 
astronomers over decades, and then report on the ``new'' exciting discoveries 
of afterglow emission at longer wavelengths which have raised more new 
questions than answered old ones.

\end{abstract}


\section{Introduction}

Gamma-ray bursts (GRBs) were first detected in 1967 with small gamma-ray
detectors onboard the Vela satellites (Klebesadel \etal\ 1973) which were 
designed to verify the nuclear test ban treaty between the USA and the USSR. 
For many years the prevailing opinion was that magnetic neutron stars (NS) 
in the galactic disk were the sources of GRBs. No flaring emission outside
the gamma-ray region could be detected, and no undisputable quiescent 
counterpart to a GRB could be established. Despite a distance ``uncertainty''
of 10 orders of magnitude, numerous theories (see a compilation in Nemiroff 
1994) were advanced to explain the source of energy in GRBs.
The measurements since 1991 of the Burst and Transient Source Experiment 
(BATSE) onboard the Compton Gamma-Ray Observatory have shown unequivocally that
GRBs {\bf are} isotropic even at the faintest intensities, and that there {\bf 
is} a distinct lack of faint bursts as compared to a homogeneous distribution.
An unprecedented wealth of additional information on each burst
could be collected, yet the GRB origin remained a mystery.

Over the previous two decades, GRB coordinates came with two mutually exclusive
properties: {\it arcmin accuracy} as provided by the interplanetary network
(Hurley 1995) or {\it fast} as provided by the BATSE Coordinate Distribution 
Network 
(BACODINE) system (Barthelmy \etal\ 1996). Only since the launch and successful
operation of the Italian/Dutch BeppoSAX satellite is it possible to obtain
accurate GRB positions in reasonably short time (few hours) which allow
quick follow-up observations (Heise \etal\ 1998).
The discovery of X-ray afterglow emission with the BeppoSAX satellite 
and related optical and radio transients has
given a dramatic boost to both observations and theoretical investigations
of GRBs over the last few months.
At the present time (late 1997), our knowledge is evolving extremely rapidly. 
Thus, it may not be surprising that the content of this review has been
expanded considerably as compared to the oral version given in May 1997.
As in the talk, I will not cover Soft Gamma Repeaters, reviews of which can
be found in Kulkarni (1998) and Smith (1998).

\section{Basic facts}

About once per day, the most sensitive gamma-ray instruments detect a short
burst of high-energy radiation from an unpredictable location in the sky.
Most of its power is radiated in the 100--500 keV range, but photons up to
18 GeV or down to a few keV have also been registered.
The bursts have durations of typically 0.1--10 sec, and the rise to
maximum intensity can occur within fractions of a millisecond. During these
short times GRBs are the brightest objects on the X-ray/$\gamma$-ray sky.

\begin{figure}[th]
      \vbox{\psfig{figure=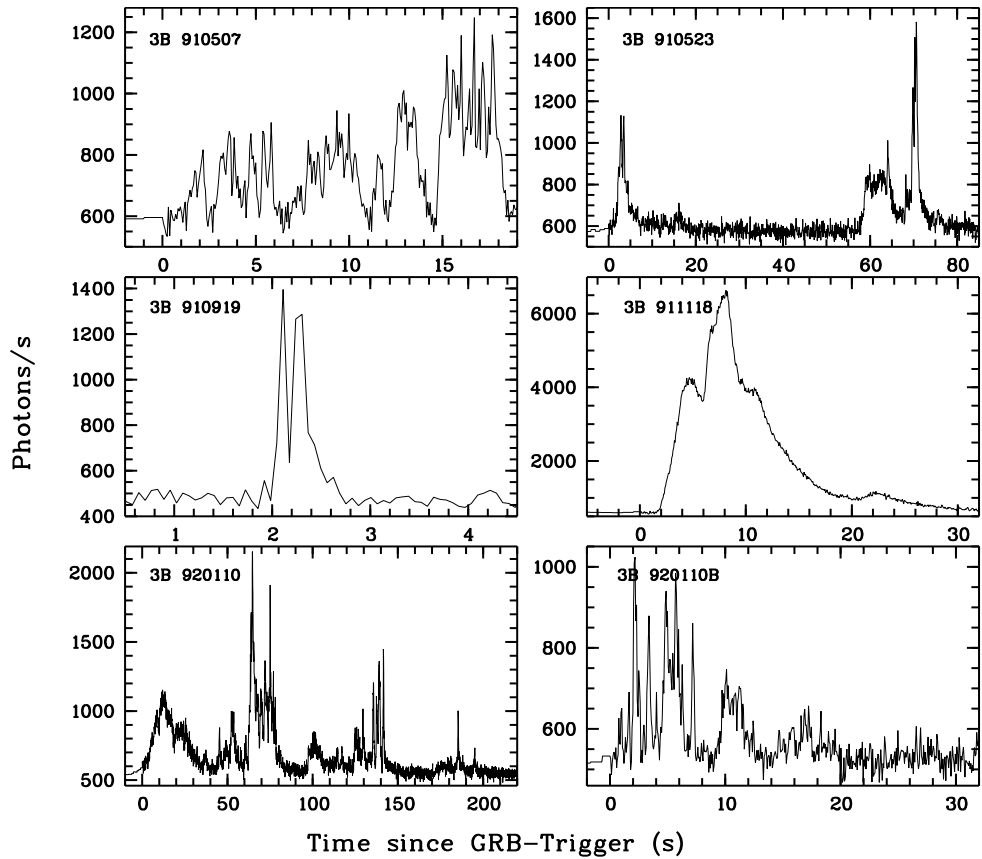,width=13.cm,%
          bbllx=1.9cm,bblly=10.8cm,bburx=18.1cm,bbury=25.cm,clip=}}\par
      \caption[grblc]{Examples for GRB light curves as seen with BATSE: 
          the photon count rate is plotted over time with a temporal resolution
          of 64 ms. Noteworthy are the diversity of structures within the
          bursts as well as their durations.
              }
      \label{lc}
 \end{figure}

The majority of GRBs has a rather complex temporal structure (Fig. \ref{lc}): 
in particular their variability time scale is significantly shorter than 
the duration (Meegan \etal\ 1996). The typical ratio between duration 
and the length of intensity peaks within a GRB is about 100. Note that the
GRB durations scatter over six orders of magnitude (1 msec to 1 ksec), 
and given additional
spectral variations from burst to burst and even within bursts, one is faced
with a confusing diversity of the ``simply measurable quantities''.

   \begin{figure}
      \vbox{\psfig{figure=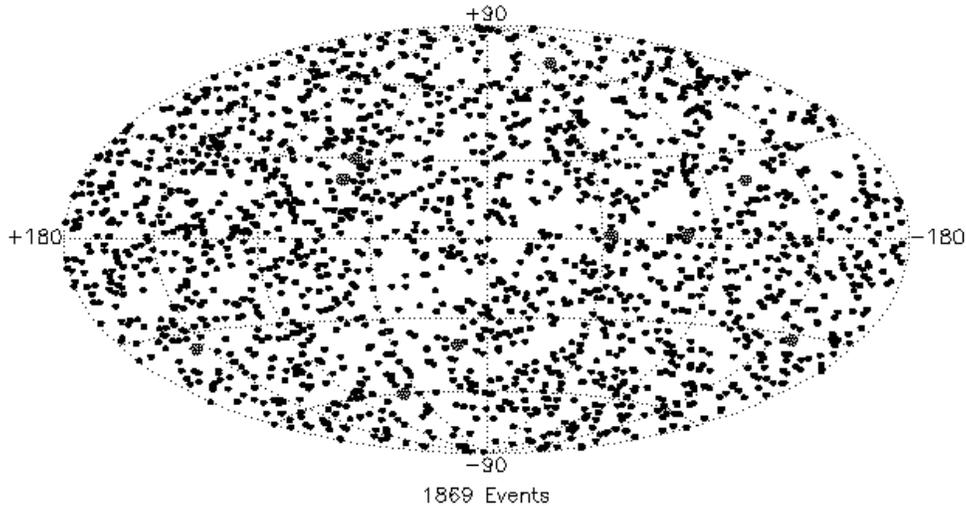,width=13.1cm,%
          bbllx=1.3cm,bblly=9.6cm,bburx=19.70cm,bbury=19.3cm,clip=}}\par
       \vspace{-0.4cm}
      \caption[grbmap]{Sky distribution of all GRBs detected with the BATSE
         instrument until July 10, 1997 (the galactic center is in the middle).
        (Adapted from the BATSE Web-page
          http:/$\!$/www.batse.msfc.nasa.gov/data/grb/skymap/)
              }
         \label{grb_map}
         \vspace{-0.2cm}
    \end{figure}

The most direct hint for the spatial distribution of GRB sources comes from the
distribution on the sky combined with the observed intensity distribution.
The latter is a folding of the unknown radial 
with the unknown luminosity distribution. In an Euclidean space, a homogeneous
source distribution will display an intensity distribution
of N($>$S)$\propto$S$^{-3/2}$. This is seen for bright bursts, but one 
observes an increasing deficit of bursts at faint intensities. Combined with 
the complete isotropy on the sky (Fig. \ref{grb_map}) this implies a
GRB distribution which is centered at the Earth and has a 
decreasing source density with increasing distance.
Four different distance scales are possible in principle:
\begin{enumerate}
\item Oort's comet cloud is the nearest population which obeys these 
characteristics. Collisions among
comets (White 1993) or between comets and primordial black holes 
(Bickert \& Greiner 1993) have been proposed as possible scenarios. 
However, the Oort cloud radius is a sizeable fraction of the
distance to nearby stars, and the expected  significant
clustering of GRBs around the nearest stars is not observed.
\item A distribution of sources along the galactic plane would
appear to be isotropic if the sampling distance is smaller than the scale
height of the disk distribution. Such distribution was very popular in the
pre-BATSE era, and magnetic NS were the favourite sources of
GRBs with typical luminosities of 10$^{38}$ erg/s.
Both, the observed quadrupole moment of the GRB distribution as well as the
fact that we see the ``faint end'' 
clearly invalidate such a distribution (Briggs \etal\ 1996).
\item An extended halo around the Galaxy has been a
viable alternative for galactic GRB distributions. Isotropy requires
a distance $\geq$100 kpc while the lack of GRB clustering around
M31 constrains the outer radius of the halo to less than $\approx$300 kpc.
However, the original scenario of high-velocity NSs
(Shklovski \& Mitrofanov 1985) escaping from the galactic disk needs
additional ad hoc assumptions to be in line with observations
(Hartmann \etal\ 1990, Greiner 1991, Podsiadlowski \etal\ 1995).
\item If GRBs are at cosmological distances, their isotropy on the sky is a 
natural consequence. Originally proposed very soon after the discovery of 
GRBs (Usov \& Chibisov 1975), this option became popular only after the
suggested collision of compact objects as a viable mechanism (Paczynski 1991).
With typical distances of z$\approx$1 the corresponding luminosities
are of the order of 10$^{51}$ erg/s. 
\end{enumerate}

\section{Theories for GRBs at cosmological distances}

The basic scenario for the understanding of the properties of low-energy
afterglows of GRBs is the dissipative (shock) {\bf fireball model} 
(Meszaros \& Rees 1993, 1997). It assumes a very large energy deposition
inside a very small volume (constrained by causality and the variability
timescales of GRBs to be of order 100 km or smaller) which leads to 
characteristic photon energy densities
which produce an optically thick, highly super-Eddington 
$\gamma$e$^{\pm}$ fireball. The fireball initially is thermal and converts
most of its radiation energy into kinetic energy, i.e. bulk motion of a
relativistically expanding blast wave (Lorentz factor $\Gamma \sim$ 
10$^{2-3}$ required to avoid degradation of the GeV photons by photon-photon 
interactions). The kinetic energy is tapped by shocks as the most likely
dissipation mechanism, and these shocks should probably occur after the
fireball became optically thin, as suggested by the observed
non-thermal GRB spectra.

Two types of fireball models can be distinguished which involve different 
explanations for the duration and variability of the GRB. In one type
(called external shocks; Meszaros \& Rees 1993) the shocks are caused 
by the interaction (collision) 
of the fireball ejecta with the surrounding medium. The typical duration 
of a GRB is then given by the Doppler delayed arrival times of the emission
from the two boundaries of the ejecta shell, or from the delay between
different surface elements within the light cone. Detailed investigations
have shown (Fenimore \etal\ 1996, Sari \& Piran 1997) that it is difficult
to produce the variety of temporal structures observed in GRBs.
The other type of model (called internal shocks;  Rees \& Meszaros 1994,
Paczynski \& Xu 1994) relates the shocks to inhomogeneities within the 
relativistic outflow, e.g. catching up of faster portions with slower
portions of the flow. The duration of these shocks is likely to be given by
the intrinsic duration of the energy release, while any intrinsic variability
(of arbitrary scale) caused by the central engine will be responsible for the
GRB time history.

If GRBs occur e.g. inside galaxies where the external medium has an 
appreciable 
density, one would expect internal shock bursts to be followed by external 
shock bursts/emission, which has twofold relevance: (1) External shocks
could be related to the observed delayed GeV emission (Meszaros \& Rees 1994),
and (2) the external shocks could radiate at lower energies and longer 
timescales  as compared to internal shocks, and thus are thought to produce
the afterglow emission. 

The afterglow emission at longer wavelengths thus is most probably 
due to synchrotron or Inverse Compton cooling from a decelerating
relativistic shell. As the
Lorentz factor of the shell decreases, the typical synchrotron frequency also
decreases, causing a delay of emission towards longer wavelengths.
Several variants of this basic scenario have been proposed including
adiabatic versus radiative hydrodynamics, fast versus slow cooling of the
electrons, and synchrotron versus synchrotron self-Compton emission.
Not all combinations are possible, and so far there is no single model
that fits all results  of afterglow observations.
Some versions of the external shock models
predict a relation between the energy spectral slope $\alpha$ (power law 
photon index) of the afterglow emission and the slope of the intensity decay
$\delta$ (I $\propto$ t$^\delta$; $\delta$ = 3/2 ($\alpha$+1)).

The generic nature of the fireball model is based on the fact that
the detailed nature of the primary energy mechanism is largely undetermined,
i.e. it can be reconciled with most of the proposed scenarios,  such as
a binary compact object merger (Paczynski 1986), 
a failed supernova (Woosley 1992),
a young highly magnetic pulsar (Usov 1992),
or a hypernova (Paczynski 1998).

\section{Rapid and accurate burst localisation}

\Ni{\bf BeppoSAX bringing the break-through:} 
With the launch of BeppoSAX, the combination of two new features allowed 
the exciting new 
discovery of X-ray afterglows, and optical/radio transients: first, the
combination of a GRB monitor (Frontera \etal\ 1997c) with
two Wide Field Cameras (WFC; Jager \etal\ 1997) with a 40\degs\ field of view
each which can localize about 10/yr 
at the few arcmin level (in't Zand \etal\ 1997); second, the ability to point
within hours after the discovery to the GRB location has enabled 
the detection of fading X-ray emission of a GRB 
for the first time (Piro \etal\ 1998a).

\Ni{\bf RXTE/PCA scanning of BATSE GRB positions:}
The discovery of long-lasting X-ray afterglows has led to the establishment
of a procedure to scan the smallest BATSE GRB error circles with the
one degree field of view (FOV; collimated) PCA detector on RXTE. Two afterglows
have been seen so far.

\Ni{\bf RXTE/ASM locations:}
The all-sky monitor (ASM) on RXTE observes the sky with its three cameras
(FOV of 6\degs$\times$90\degs) in 90 s stationary exposures 
followed by an instrument rotation of 6\degs. The sky coverage
implies an X-ray afterglow detection rate of 8-10 GRBs per year, and
several GRBs have been localized in 1997 (Smith \etal\ 1998).

\Ni{\bf Interplanetary network (IPN):}
The Ulysses spacecraft has been operational for more than 7 years now,
but has been the only interplanetary mission with a GRB detector since the
failure of the Mars missions. Thus, GRB locations as provided by BATSE
could be reduced only in one dimension to a few arcmin width (Hurley 1995). 
Over the last months, the $\gamma$-ray spectrometer on the Near Earth Asteroid 
Rendezvous (NEAR) spacecraft has been reconfigured to allow measurements
of GRBs (the spectrometer was not originally planned to begin working until
NEAR reached Eros in Feb. 1999). The first detection of a GRB occurred on
Sep. 15, 1997 and several more GRBs have been detected since then several
of which have been seen also by the BATSE, Ulysses or Wind spacecrafts. 
Thus, the new capability  of NEAR adds a new dimension to the IPN and
enables to obtain locations of moderate and strong GRBs with arcmin accuracy.

\Ni{\bf The future:}
The future seems bright for detection of X-ray afterglows of GRBs: 
after the launch of the original HETE mission it is presently being
rebuilt with soft X-ray cameras replacing the UV cameras. The all-sky monitor 
MOXE onboard the Russian SRG mission will monitor the sky nearly continuously
with a spatial resolution down to 1 arcmin. And the German ABRIXAS satellite
will scan the sky with its seven identical 40 arcmin FOV telescopes on great 
circles, similar to ROSAT (Tr\"umper 1983). All these instruments are scheduled
for launch in 1999, so a substantial improvement in the rate of GRB X-ray 
afterglow detections can be expected in the near future.

\section{Fading Counterparts}

\subsection{X-ray afterglows}

\subsubsection{BeppoSAX narrow-field instruments pointings}

The ability to rapidly point  the narrow-field instruments (NFI)
towards GRB positions has enabled the discovery of long-lasting X-ray
emission which decayed according to a $\approx$t$^{-1}$ power law.
While neither the power law  nor the slope are unique, the
oc\-cur\-rence within a 3\amin\ GRB error box right after the burst 
and the correlation with 
similarly decaying optical transients is convincing evidence for its 
relationship to the GRB.

With only one exception so far, BeppoSAX NFI pointings have been
performed for all GRBs localized with the WFC (see Tabs. 
\ref{locgrb}, \ref{nfi}). With the exception of GRB 970111 X-ray afterglow
emission has been clearly detected in all pointings which occurred within less 
than 2 days of the GRB; there is recent evidence for a possible X-ray afterglow
also of GRB 970111: an X-ray source is detected in the first half of the
observation (Feroci \etal\ 1998).
Unfortunately, not in all cases were detections made
in both detectors (LECS:  0.1--10 keV;  MECS: 2--10 keV),
and in most cases the detected LECS rates are too small to allow any
detailed spectral investigations. However, three important results
could be established (e.g. Frontera \etal\ 1998b): 
(i) the afterglow X-ray spectrum seemingly is
well fitted by an absorbed power law model, but not with thermal models
(bremsstrahlung or blackbody), (ii) there are no obvious changes in 
the shape of the X-ray spectrum along the intensity decay, and
(iii) within the accuracy of measurements indeed the intensity decay
slope $\delta$ is related to the spectral slope of the energy spectrum
of the GRB tail and afterglow.

\begin{table}[ht]
\caption{GRBs localized by BeppoSAX and RXTE}
\begin{tabular}{|l|c|c|c|c|c|c|c|}
\hline
    \rule[-3mm]{0mm}{6.5mm} GRB & GRB X-ray position & Error & Instrument 
      & IPN$^{a}$ & XA$^{a}$  & OT$^{a}$ & Ref.$^{b}$\\
\hline
   960720 & 17\h30\m37\s  +49\degs 05\farcm8      &  3\amin   & SAX/WFC & 
       n & n & n & 1, 3 \\
   970111 & 15\h28\m15\s  +19\degs 36\farcm3      &  3\amin   & SAX/WFC & 
       y & y? & n & 2, 3 \\
   970228 & 05\h01\m57\s  +11\degs 46\farcm4      &  3\amin   & SAX/WFC & 
       y & y & y & 4 \\
   970402 & 14\h50\m16\s  \,--\,69\degs 19\farcm9 &  3\amin   & SAX/WFC & 
       n & y & n & 5 \\
   970508 & 06\h53\m28\s  +79\degs 17\farcm4      &  3\amin   & SAX/WFC & 
       n & y & y & 6, 7\\
   970616 & 01\h18\m57\s  \,--\,05\degs 28\farcm0 & 40\amin x2\amin& XTE/Uly & 
       y & y & n & 8, 9\\
   970815 & 16\h08\m43\s  +81\degs 30\farcm6      & 6\amin x3\amin & XTE ASM & 
       n & y & n & 10\\
   970828 & 18\h08\m29\s  +59\degs 18\farcm0      & 2\farcm5x1\amin & XTE ASM &
       y & y & n & 11--13\\
   971024 & 18\h24\m51\s  +49\degs 28\farcm9      & 9\farcm0x1\amin & XTE ASM &
       n & y & n & $\!\!$priv. comm.$\!\!$\\
   971214 & 11\h56\m30\s  +65\degs 12\farcm0      & 4\amin     & SAX/WFC & 
       y & y & y & 14, 15\\
   971227 & 12\h57\m35\s  +59\degs 15\farcm4      & 8\amin     & SAX/WFC & 
       n & y?& y? &  16\\
\hline
\end{tabular}

\noindent{\Ni\small
   $^{(a)}$ IPN = interplanetary network detection; 
            XA = X-ray afterglow; OT = optical transient. \\
   $^{(b)}$ (1) Piro \etal\ 1996,        (2) Costa \etal\ 1997a,       
            (3) in't Zand \etal\ 1997, 
            (4) Costa \etal\ 1997b,      (5) Feroci \etal\ 1997,      
            (6) Costa \etal\ 1997d,      (7) Heise \etal\ 1997a, 
            (8) Marshall \etal\ 1997,    (9) Hurley \etal\ 1997a,
           (10) Smith \etal\ 1997a,     (11) Remillard \etal\ 1997,
           (12) Smith \etal\ 1997b,     (13) Hurley \etal\ 1997b,
           (14) Heise \etal\ 1997b,     (15) Kippen \etal\ 1997,
           (16) Coletta \etal\ 1997}
\label{locgrb}
\end{table}

   \begin{table}[htb]
     \vspace*{-0.1cm}
     \caption{BeppoSAX NFI pointed observations of GRBs}
     \begin{tabular}{|c|c|c|c|c|c|}
     \noalign{\smallskip}
     \hline
           GRB & Exposure & Delay & $N_{\rm X}$$^{(a)}$ & 
             \rule[-3mm]{0mm}{7mm}X-ray afterglow flux &   Ref.$^{(b)}$ \\
               & (sec; MECS) &     &          & (erg/cm$^2$/s; 2--10 keV) &  \\
     \hline
    ~960720~ & 56\,000 & 43 days & 1 & -- & 1 \\
      970111 & 52\,000 & 16 hrs  & 2 &    2$\times$10$^{-13}$ & 2, 3 \\
      970228 & 15\,000 & ~~8 hrs   & 1 &    3$\times$10$^{-12}$ & 4 \\
             & 16\,270 & 87 hrs  & 1 &    2$\times$10$^{-13}$ & 4 \\
      970402 & 25\,000 & ~~8 hrs   & 1 &    6$\times$10$^{-13}$ & 5 \\
             & 50\,000 & 41 hrs  & 0 & $<$2$\times$10$^{-13}$~~ & 5 \\
      970508 & 25\,000 & 5.7 hrs & 1 &    6$\times$10$^{-13}$ & 6, 7 \\
             & 24\,000 & 6.6 hrs & 1 &    4$\times$10$^{-13}$ & 8 \\
             & 12\,000 & 99 hrs  & 1 &  2.4$\times$10$^{-13}$~~ & 8 \\
             & 73\,000 & 137 hrs & 1 &    5$\times$10$^{-14}$ & 8 \\
      971214 &101\,200~ & 6.5 hrs & 1 &    4$\times$10$^{-13}$ & 8, 9 \\
      971227 & 14\,200 & 14 hrs  & 2 &    3$\times$10$^{-13}$ & 8, 10 \\
      \hline
      \noalign{\smallskip}
   \end{tabular}
   \label{nfi}

\noindent{\Ni 
   $^{(a)}$ $N_{\rm X}$ is the number of X-ray sources found inside the
           GRB error box. \\
   $^{(b)}$ (1) Piro \etal\ 1998a,        (2) Butler \etal\ 1997,
            (3) Feroci \etal\ 1998,       (4) Costa \etal\ 1997c,      
            (5) Piro \etal\ 1997a,        (6) Piro \etal\ 1997b,
            (7) Piro \etal\ 1998b,
            (8) priv. comm.,
            (9) Antonelli \etal\ 1997,   (10) Piro \etal\ 1997c.}
   \vspace*{-0.2cm}
   \end{table}

\subsubsection{Quick ASCA and ROSAT follow-up observations of GRBs}

Motivated by the occurrence of a few very long lasting GRBs and the
detection of distinct spectral softening over the burst duration a number of
attempts have been made in the past
to observe well-localized GRB locations with ROSAT and ASCA as quickly as 
possible after the GRB event, in the hope to find the ``smoking gun''.
To this end, the GRB had not only to be localized quickly, 
but the GRB location also had to be within the
ROSAT/ASCA observing windows ($\approx$30--40\% of the sky at any moment). 
With the quick and accurate location capabilities of BeppoSAX, these 
attempts have been intensified and since then practically every observable
location of a well-localized GRB has been observed with ROSAT and ASCA.
The fastest response with ASCA (ROSAT) so far is 1.2 (5) days,
which is near the minimum
possible time achievable due to the various scheduling constraints.
The primary goals are to determine accurate positions at the 10\asec\ level
(ROSAT), to measure the afterglow X-ray spectrum (ASCA) and to follow the
intensity decay curve beyond the abilities of BeppoSAX (ROSAT/ASCA).
Tab. \ref{toos} lists
the GRBs which have been observed as TOO together with the time delay
between the GRB and the ROSAT/ASCA observation.

   \begin{table}[htb]
     \vspace{-0.1cm}
     \caption{Rapid ROSAT and ASCA target-of-opportunity observations 
          (TOO) towards GRB locations}
     \begin{tabular}{|c|r|c|c|l|r|c|c|l|}
      \noalign{\smallskip}
      \hline
           GRB & \multicolumn{4}{c|}{\rule[-3mm]{0mm}{7mm}ROSAT}
           & \multicolumn{4}{|c|}{ASCA} \\
      \cline{2-9}
      & $\!\!$Exposure$\!\!$~ & \rule[-3mm]{0mm}{6.5mm} Delay & N$_{\rm X}$ 
                                                          & Ref.$^{a)}$ &
           $\!\!$Exposure$\!\!$ & Delay & N$_{\rm X}$ & Ref.$^{a)}$ \\
       &  (sec)~~             &       &             &            &
          (sec)~~             &       &             &           \\
      \hline
   920501     & 2\,748~~ & 18 days  & 1 & 1    & 40\,000~  & 3 yrs & 1 & 2 \\
   920711     & 2\,432~~ & 28 weeks & 0 & 3, 4 &  --~~  & -- & -- & -- \\
   $\!\!$930704/940301$\!\!$& 3156/1385 & ~4 weeks & 25 & 5 &
                 --~~  & -- & -- & -- \\
   960720     &  6\,960~~ & 42 days & 1 & 6, 7 & 45\,630~ & 60 days & 0& 8\\
              & 2\,791~~ & 24 weeks  & 1  &  4 & --~~   & -- & -- & -- \\
   961027--29 & 2065/2499 & 13 weeks & 54 & 4 & --~~ & -- & -- & -- \\
   970111     & 1\,198~~ & ~5 days   & 0  & 9 & --~~ & -- & -- & -- \\
              &    777~~ & 38 days   & 0  &  4 & --~~  & -- & -- & -- \\
   970228     & 34\,280~~ & $\!\!$11--14 days$\!\!$ & 1 & 10, 11 &  
                19\,930~ & 7 days & 1 & 12 \\
   970402     & --~~ & --    & -- & -- & 39\,100~ & 2.8 days & 0 & 13 \\
   970616     & 21\,950~~ & 7--9 days & 2 & 14 & 50\,000~~&3.5 days & 2 & 15 \\
   970815     & 17\,115~~ & 5--7 days & 1 & 16 & 46\,000~~&3.2 days & 0 & 17 \\
   970828     & 61\,275~~ &6--8 days & 1& 18& 38\,950~~& 1.2 days & 1& 19, 20\\
   971024     & 14\,898~~ &  6 days   & 0  & 4 & --~~ & -- & -- & -- \\
      \hline
      \noalign{\smallskip}
   \end{tabular}
   \label{toos}
   \vspace*{-0.1cm}

   \noindent $^{a)}$ (1) Hurley \etal\ (1996), (2) Murakami (1996a),
                     (3) PI: Hurley,           (4) Greiner (unpubl.),         
                     (5) Greiner \etal\ (1997a), 
                     (6) Greiner \etal\ (1996),    (7) Greiner \& Heise (1997),
                     (8) Murakami \etal\ (1996b),  (9) Frontera \etal\ (1997a),
                    (10) Frontera \etal\ (1997b), (11) Frontera \etal\ (1998a),
                    (12) Yoshida \etal\ (1997),  (13) Murakami \etal\ (1998),
                    (14) Greiner \etal\ (1997b), (15) Murakami \etal\ (1997a),
                    (16) Greiner (1997),         (17) Murakami \etal\ (1997b),
                    (18) Greiner \etal\ (1997c),
                    (19) Murakami \etal\ (1997c), (20) Murakami \etal\ (1997d).
   \vspace*{-0.25cm}
   \end{table}

\subsubsection{Deep ROSAT/ASCA observations of GRB error boxes}

Over the last seven years nearly a dozen GRB error boxes were observed with the
ROSAT and ASCA satellites for up to 80 ksec exposure time (see Tab. 1 in
Greiner (1998) for a complete listing of ROSAT pointings), 
thus improving considerably the sensitivity limits obtained
with earlier X-ray observations at soft energies. X-ray sources have been 
found inside the error boxes of some of these GRB. 
While originally the discovery of
a quiescent X-ray source inside a small GRB error box was considered as
probable evidence for an association of a GRB with a quiescent counterpart,
the continuing discovery of further X-ray sources and in particular the
detection of more than one X-ray source even in small GRB error boxes 
 makes this association doubtful. Also, the optical
identification of these X-ray sources, though not yet completely established 
in all cases, does not provide evidence for unusual objects.
The present knowledge of properties of fading X-ray afterglows also argues 
against an association of these X-ray sources with the GRB, and makes the lack
of ``success'' of these deep pointings understandable.

An estimate of the chance probability for the occurrence of a quiescent
X-ray source inside a small GRB error box depends on how the 
question is asked in detail (see Hurley \etal\ 1996 for various possibilities).
However,  at the low sensitivity limits reached, 
the number density of X-ray sources is already remarkably high. From 
the results of many deep pointed observations and combined with the
very deep Lockman hole observations of ROSAT, an improved log\,N--log\,S
distribution of X-ray sources has been derived (Hasinger 1997) which
gives 100--700 X-ray sources per 1\,$\Box$\grad\, 
at the level of 10$^{-14}$...10$^{-15}$ erg/cm$^2$/s.
Thus, the probability for a chance coincidence of a quiescent, soft X-ray 
source with a GRB location is 25\%--100\% for a 10 arcmin$^2$ size error box.

 \begin{figure}[th]
      \vbox{\psfig{figure=grb970228_galama.ps,width=6.3cm,%
          bbllx=1.5cm,bblly=3.6cm,bburx=16.7cm,bbury=15.cm,clip=}}\par
      \vspace{-4.7cm}\hspace*{6.6cm}
      \vbox{\psfig{figure=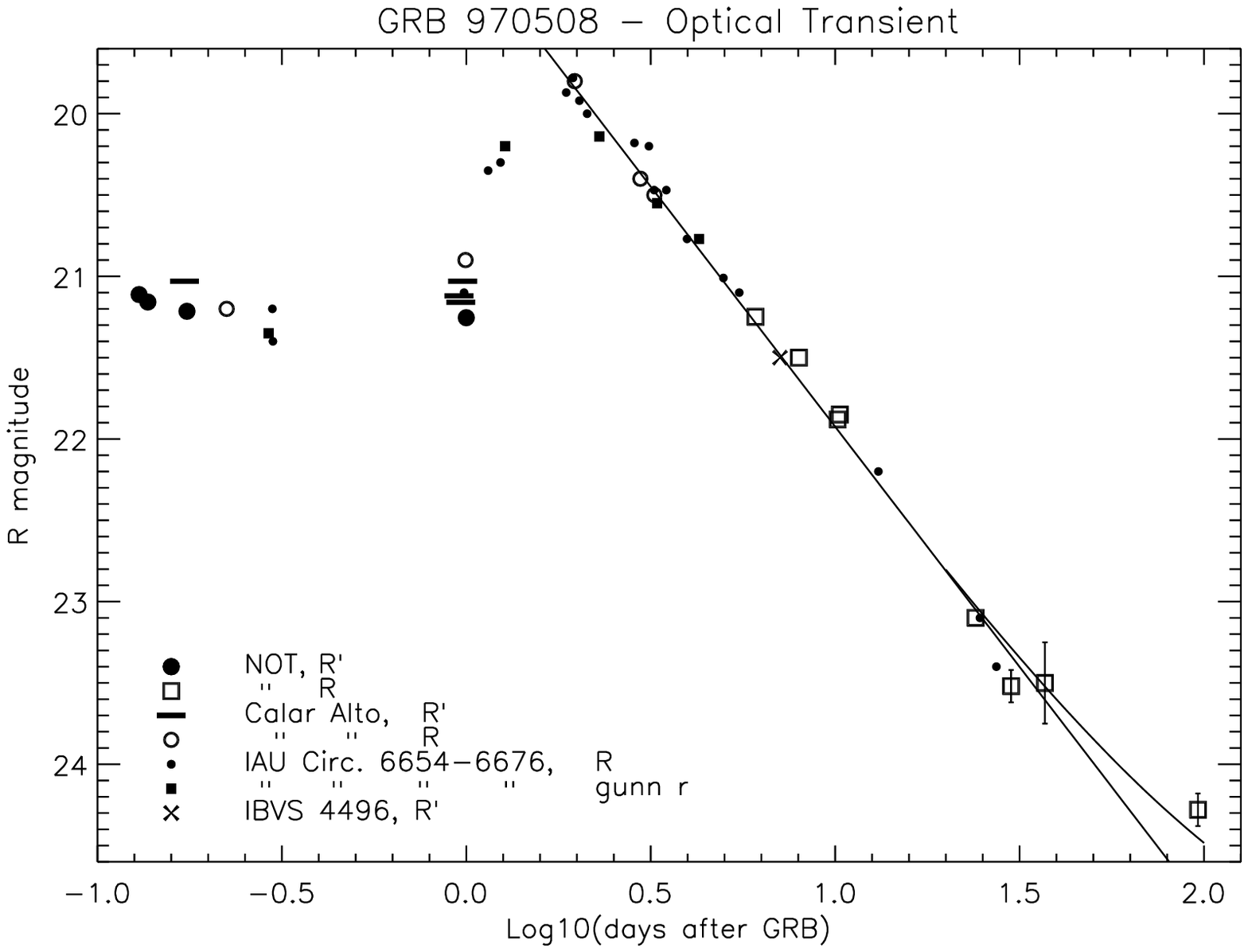,width=6.5cm,%
          bbllx=2.8cm,bblly=12.9cm,bburx=19.1cm,bbury=24.7cm,clip=}}\par
      \vspace{-0.3cm}
      \caption[olc]{R band lightcurves of GRB 970228 (left; Galama \etal\ 
            1998) and GRB 970508 (right; Pedersen \etal\ 1998).
            The intensity decay is well described by a power law of 
            T$^{-1.1}$ and T$^{-1.2}$, respectively. Note the plateau and 
            initial rise during the first 2 days in GRB 970508.
              }
      \label{olc}
      \vspace{-0.2cm}
 \end{figure}

\subsection{Optical transients}

Optical transients have been discovered so far for three well-localized GRBs
with a fourth one (GRB 971227) being disputed. Only a short summary is given
here, and much more details can be found on the GCN pages
(http:/$\!$/gcn.gsfc.nasa.gov/gcn/gcn\_main.html) or on a nearly systematic
data collection at http:/$\!$/www.aip.de:8080/\~\,jcg/grbgen.html and 
references/links therein:

\noindent {\bf GRB 970228:}
The first optical transient (OT) associated with a GRB was discovered
on deep images taken 21 hrs after the GRB as a 20 mag object fading by
more than 3 magnitudes during the first 4 days (van Paradijs \etal\ 1997).
The optical power law intensity decay of GRB 970228 is unbroken until the
latest observations (8 months after the burst; Fruchter \etal\ 1998), thus 
arguing for an early onset of the adiabatic expansion regime of the fireball 
and indicating that the nonrelativistic regime still has not been reached.
The OT is surrounded by fuzzy emission, most probably a galaxy, and this
association first suggested that GRBs lie at cosmological distance.
The proper motion of the OT reported by Caraveo \etal\ (1997) has been
disputed (e.g. Sahu \etal\ 1997).

\noindent {\bf GRB 970508:}
The spectrum of the OT of GRB 970508 has shown a 
pattern of absorption lines which are interpreted as strongly redshifted 
(z=0.835) Fe and Mg lines. If the spectrum of the optical counterpart is
featureless (according to theoretical prediction) it may
have been absorbed by a faint galaxy on the line of sight, implying a 
distance to this GRB of  z$>$0.835 (Metzger \etal\ 1997), thus providing 
the first direct distance determination of a GRB.
The ``unusual'' optical lightcurve, i.e. a rise after a two day long plateau,
followed by the power law decay similar to that in GRB 970228, has been
interpreted as emission from a hot cloud where the spectrum peaks well above
optical frequencies and gradually shifts down during its expansion
(Katz 1994, Meszaros 1998).

\noindent {\bf GRB 971214:} The optical transient of GRB 971214 was rather
faint (I$\approx$21 mag) right from the first observations 10 hrs after the
burst. The point-like OT decayed quickly to below 25th mag within the first
three days. Optical spectra obtained on Dec. 17 and 28, 1997 revealed
extended emission features, but did not allow a unique redshift
determination yet (Kulkarni \etal\ 1998).

\subsection{Radio transient(s)}

Despite intensive monitoring of GRBs from a few hours to several months
after its occurrence (Frail 1998) radio emission
has been securely detected only from one burst - GRB 970508.
First detected with the VLA on May 13, the radio emission rose and fell 
several times over the next weeks, and the shape of the radio spectrum also
changed (Frail \etal\ 1997, Taylor \etal\ 1997). Nearly at the same time, 
Goodman (1997) predicted that irregularities in an extremely tenuous
ISM could cause fluctuations in the radio intensity if the source size is
small enough. Scattering produces multiple images of the source, and 
interference between the multiple images may produce a diffraction pattern
(perpendicular to the line of sight), leading to a strong flux variation
on a time scale of $\approx$3 hrs as the observer (with Earth's and Sun's 
peculiar motion of $\approx$30 km/s) moves across the pattern.

%

Comparison of the expansion rate of the fireball model (see above) suggests
that on a time scale of weeks the apparent fireball size grows to the 
maximum size for which diffractive scintillation is possible 
(Waxman \etal\ 1998). This is 
exactly what is observed: after about 2 months the radio ``twinkling''
of GRB 970508 slowly decreased. Thus, once a cosmological distance is 
adopted for GRB 970508 (based on the absorption line systems in the optical 
spectrum), the radio measurements imply an apparent 
{size\,of\,$\approx$10$^{17}$\,cm} about 1 months after the GRB and an expansion
velocity near the speed of light.

\subsection{Scaling of basic parameters}

Relativistic fireball models predict that basic parameters of
GRB afterglows are scaled to each other: total energy, initial Lorentz factor,
surrounding gas density and distance. However, based on the measured fluxes
from radio to $\gamma$-rays, all afterglows observed
so far have exhibited completely unscaled behaviour in two ways. 
(a) X-ray afterglows in a rather tight range of intensity are detected from 
GRBs with a wide range of $\gamma$-ray parameters (e.g. peak flux, duration,
temporal structure), and (b) optical and radio intensities are seemingly
neither connected to peak $\gamma$-ray flux nor X-ray afterglow flux.
In particular, the lack of optical transients in some GRBs with strong X-ray 
afterglow emission is hard to understand in terms of intrinsic properties 
related to the fireball model, since the optical emission in any case should be
less beamed than the X-rays, and thus be more frequent. However, the external
medium could play a significant role, i.e. possible absorption by 
interstellar gas or dust in the host (Jenkins 1997, Paczynski 1998).
If absorption occurs in the host at cosmological distances, the measured
low-energy cut-off in the X-ray spectrum will be lower than the intrinsic 
absorption by a factor of (1+z). Similarly, the observed optical emission has 
a wavelength shorter by a factor (1+z) thus suffering stronger extinction at a
given absorbing column. As an example, at z=1 a column of 10$^{22}$ cm$^{-2}$
would correspond to an effective X-ray absorbing column of
1.5$\times$10$^{21}$ cm$^{-2}$ and thus only marginally 
be distinguishable in the X-ray spectrum from galactic foreground absorption, 
but the observed I band emission actually is absorbed by the corresponding 
$A_{\rm V}$=7 in the host. Therefore, even small changes in the column within
the GRB host can drastically reduce the optical flux while leaving the
low-energy cut-off of the X-ray emission nearly unaffected. 

While the above effect is important for the IR to X-ray range, the primary 
$\gamma$-rays as well as the radio emission are not affected. Thus, if more 
radio transients are discovered in the next months this should allow us to 
test directly the prediction of scaling among different GRBs by comparing their
$F_{\rm \gamma}/F_{\rm radio}$, avoiding the uncertainties
imposed by the possible intrinsic host absorption.

\subsection{Consequences for the sources of GRBs}

If the explanation on the lack of optical transients relative to X-ray
afterglows turns out to be correct, then GRBs would be linked
to dense star forming regions, and thus possibly to a population of
massive stars. Since the age of a massive star is not more than a few million
years, it explodes within its star forming region. In the proposed
hypernova scenario (Paczynski 1998), a massive and rapidly spinning star
may release $\sim$10$^{54}$ erg of kinetic energy extracted from the
rotational energy of the black hole (Blandford \& Znajek 1977).
Only a small fraction of this kinetic energy is in the debris ejected 
with the largest Lorentz factors (which are required to generate the 
$\gamma$-rays), while most of the ejecta is sub-relativistic (speed of c/3 for 
a 10 \msun\ object). Thus, when the fireball is slowed down by the
ambient medium, the slower moving ejecta gradually catch up and provide
a long lasting energy supply to the afterglow (longer than in the standard 
fireball model, but see Rees \& Meszaros 1998).

If the GRB rate follows the massive star formation rate (with redshift) 
then an immediate
consequence is an increase of the distance scale to GRBs (Sahu \etal\ 1997,
Wijers \etal\ 1998).
The increase in the comoving GRB rate with z compensates various redshift
effects which are responsible for the roll-over in the counts. The higher
energy per burst comes along with a reduced GRB rate.
If the argument is turned around, then the redshift distribution of GRBs
can, given enough statistics, be used as an independent test for the
cosmic history of the star formation rate.

Alternatively, the gravitational collapse of supermassive stars 
($M$ $\ge$ 5$\times$10$^4$ \msun) has been proposed as a cosmological source
of GRBs (Fuller \& Shi 1998, Zinnecker 1998). Since the formation, evolution 
and collapse
of supermassive stars could be pregalactic, no galaxy hosts are required.

\section{Conclusion}

The discovery of long-wavelength afterglow emission lasting for days to
months constitutes a turning point in GRB research.
However, despite these exciting new data most of the basic questions
remain unanswered. While a cosmological distance scale seems to 
be generally accepted, the nature of the GRB host remains open. Furthermore,
the complexity of GRB time histories, duration and spectra as well as 
spectral evolution during the bursts are not easily explained in the various
variants of the fireball model which is widely accepted as the main scenario 
for the production of the $\gamma$-ray burst emission.
Finally, the origin of the GRB energy source remains a mystery.
Possibly, knowledge of the history of the early cosmic evolution is required
to understand the origin of GRBs. Thus, despite (or due to) the discovery of 
flaring counterparts at longer wavelengths GRBs remain an exciting field of 
research in the foreseeable future.

\acknowledgements
I very much appreciate substantial travel support from the conference 
organizers. I made use of the electronic version of the GRB bibliography
kindly provided by K. Hurley.
The author is supported by the German Bundesministerium f\"ur Bildung, 
Wissenschaft, Forschung und Technologie (BMBF/DLR) under contract No. 
50 QQ 9602 3. 
The ROSAT project is supported by the BMBF/DLR and the Max-Planck-Society.


\end{document}